\documentclass[12pt]{iopart}
\usepackage{iopams}

 \usepackage{color}

\newcommand{\nc}{\newcommand}
\def\rr#1{(\ref{#1})}
\nc{\be}{\begin{equation}} \nc{\ee}{\end{equation}}
\nc{\ba}{\begin{array}} \nc{\ea}{\end{array}}
\nc{\bea}{\begin{eqnarray}} \nc{\eea}{\end{eqnarray}}
\nc{\ny}{\nonumber}
\nc{\ra}{\rangle} \nc{\la}{\langle}
\nc{\lk}{\left(} \nc{\rk}{\right)}

\def\sx#1{\sigma^{\rm x}_{#1}}
\def\sy#1{\sigma^{\rm y}_{#1}}
\def\sz#1{\sigma^{\rm z}_{#1}}

\def\qu{{q}}
\def\pu{{p}}

\nc\ve{{\varepsilon}}
\nc\kp{{\varkappa}}
\nc{\lm}{\lambda}
\nc\g{{\gamma}}

\nc{\IJMP}{{\it Intern. J. Mod. Phys.}}
\nc{\JStP}{{\it J. Stat. Phys.}}

\begin{document}

\title[Form-factors of the finite quantum XY-chain]{Form-factors of the finite quantum XY-chain}

\author{Nikolai Iorgov}

\address{Bogolyubov Institute for Theoretical Physics, Kiev 03680, Ukraine}
\ead{iorgov@bitp.kiev.ua}
\begin{abstract}
Explicit factorized formulas for the matrix elements (form-factors) of the spin operators $\sx{}$ and $\sy{}$ between the eigenvectors
of the Hamiltonian of the finite quantum periodic XY-chain  in a transverse field were derived.
The derivation is based on the relations between three models: the model of quantum XY-chain, Ising model on 2D lattice and
$N=2$ Baxter--Bazhanov--Stroganov $\tau^{(2)}$-model.
Due to these relations we transfer the formulas for the form-factors of the latter model
recently obtained by the use of separation of variables method to the model of quantum XY-chain.
Hopefully, the formulas for the form-factors will help in analysis of multipoint dynamic correlation functions at a finite temperature.
As an example, we re-derive the asymptotics of two-point correlation function in the disordered phase without
the use of the Toeplitz determinants and the Wiener--Hopf factorization method.
\end{abstract}

\submitto{\JPA}

\section{Introduction}

The quantum XY-chain is one of the simplest models which is rich enough from the point of view of physics
and at the same time permits strict mathematical analysis.
The study of this model was started in \cite{LSM}
where it was rewritten in terms of fermionic operators by means of the Jordan--Wigner transformation.
Now this relation is a standard mean to study different properties
(the spectrum of the Hamiltonian \cite{LSM,Katsura},
the correlation functions \cite{BarouchMcCoy,BarouchMcCoyIII,BarouchMcCoyIV,Kapiton1,Kapiton2},
the emptiness formation probability \cite{FranchiniAbanov},
the entanglement entropy \cite{Kitaev,Peschel,Its,FIK,FIKT}), quantum quenches \cite{GLFC} in XY-chain.
Although the Hamiltonian of the model is equivalent to the Hamiltonian of a free fermionic system, the spin operators
$\sx{}$ and $\sy{}$ are expressed in terms of fermionic operators in a non-local way. Thus
the study of correlation functions of such operators is a non-trivial problem. For example,
the correlation function $\la\sx{0}\sx{d}\ra$ can be written through Toeplitz determinant of size $d$
and the derivation of the asymptotics $d\to\infty$ requires \cite{BarouchMcCoy} the use of Szeg\"o theorem
and the Wiener--Hopf factorization method.

In this paper we propose an alternative way to study correlation functions of the XY-model:
we derive the formulas for the matrix elements of spin operators $\sx{}$ and $\sy{}$ between the eigenvectors
of the Hamiltonian of the finite quantum XY-chain in a transverse field. These formulas allow to obtain at least formal expression
for multipoint dynamic correlation functions at a finite temperature. For this aim, it is enough to insert the resolution
of identity operator as a sum of projectors to the eigenspaces of Hamiltonian.
Hopefully the correlation functions in terms of these sums will be more easily analyzed.
As an application of the formulas for form-factors, we re-derive the asymptotics
of  correlation function $\la\sx{0}\sx{d}\ra$ at $d\to \infty$.

The idea of derivation of form-factors of the quantum finite XY-chain is to use the relations
between three models: the model of quantum XY-chain, the Ising model on 2D lattice and
$N=2$ Baxter--Bazhanov--Stroganov (BBS) model \cite{BaxInv,BS}.
The relation between the first and the second model was observed in \cite{Suzuki}
(the relation \rr{relge} in this paper between the energies of fermionic excitations of these two models seems to be new),
the relation between the second and the third model was found in \cite{BIS,gips}.
The parameters of the models are ($h,\kp$), ($K_x,K_y$) and ($a,b$), respectively.
Due to these relations we transfer the formulas for the form-factors of  $N=2$ BBS model,
recently obtained \cite{gipst1,gipst2} by the use of separation of variables method, to the model of quantum XY-chain.
The main formulas are \rr{MEsx}, \rr{MEsy} together with \rr{ME_BL}, \rr{C}, \rr{relge}.

The separation of variables method for the quantum integrable systems (with basic example being the Toda chain)
was introduced by Sklyanin \cite{Skly1} and further developed by Kharchev and Lebedev \cite{KharLeb}.
In \cite{gips}, this method was adapted for BBS model (a $\mathbb{Z}_N$-symmetric quantum spin system) to obtain the eigenvectors of this model.
At $N=2$ and special values of parameters the BBS model reduces \cite{BIS,gips,gips_rev} to the Ising model.
The eigenvectors of transfer-matrix of Ising model obtained by the separation of variables method
allowed \cite{gipst1,gipst2} to prove the conjectural formula \cite{BL1,BL2} for the matrix elements of spin operator for
finite Ising model. This derivation had provided first proof of the formula.
A summarizing overview of the results on separation of variables for BBS model is given in \cite{gips_rev}.
It is interesting that factorized formulas for the matrix elements of spin operators exist also for
superintegrable $\mathbb{Z}_N$-symmetric chiral Potts quantum chain \cite{BaxterZN,gipstZN}.

In Sect.~2 we remind the definition of the finite quantum XY-chain in a transverse field,
its phase diagram, eigenvalues of the Hamiltonian and give general comments on
the matrix elements of spin operators between the eigenvectors of the Hamiltonian.
Sect.~3 is devoted to the description of relations between three models:
the model of quantum XY-chain, the Ising model on 2D lattice and
$N=2$ Baxter--Bazhanov--Stroganov model. Using these relations, in Sect.~4 we derive
formulas for the matrix elements (form-factors) of the spin operators $\sx{}$ and $\sy{}$ between the eigenvectors
of the Hamiltonian of the finite quantum XY-chain. In Sect.~5, these formulas are rewritten to the case of chain of infinite length.
In Sect.~6, as an application of the formulas for form-factors, we re-derive the asymptotics
\cite{BarouchMcCoy} of  correlation function
$\la\sx{0}\sx{d}\ra$ at $d\to \infty$ without the use of the Toeplitz determinants and Wiener--Hopf factorization method.

\section{Definition of the finite quantum XY-chain in a transverse field}

\subsection{The Hamiltonian and phase diagram}

The Hamiltonian of the XY-chain of length $n$ in a transverse field $h$ is \cite{LSM,Katsura}
\be\label{HamXYh}
{\cal H}=-\frac{1}{2}\sum_{k=1}^n \left(\frac{1+\kp}{2}\sx{k} \sx{k+1}\,
 +\frac{1-\kp}{2}\sy{k} \sy{k+1}\,+\,h\, \sz{k}\right)\,,
\ee
where $\sigma^i_k$ are Pauli matrices, $\kp$ is the anisotropy.
In the case $\kp=0$ we get XX-chain (isotropic case). The value $\kp=1$ corresponds to the quantum Ising chain in a transverse field.
In what follows we restrict ourselves to the case $\kp>0$, $h\ge 0$. Other signs of $\kp$ and $h$ can be obtained using
automorphisms of the algebra of Pauli matrices. Also we will suppose the periodic boundary condition
$\sigma^i_k=\sigma^i_{k+n}$. In \cite{ILff} it is shown that the formulas for the matrix elements of spin operators
obtained in Sect.~\ref{MEsection}
are also applicable for the antiperiodic boundary condition $\sx{k+n}=-\sx{k}$, $\sy{k+n}=-\sy{k}$ and $\sz{k+n}=\sz{k}$.

Now about the values of $h$. Due to the relation of XY chain with 2D Ising model, which will be discussed in the next section,
the coupling constant $h$ plays the role of a temperature-like variable. The value $h>1$ corresponds to the paramagnetic (disordered) phase.
The value $0\le h<1$ corresponds to the ferromagnetic (ordered) phase.
If $0\le h<(1-\kp^2)^{1/2}$, it is oscillatory region (because of oscillatory behavior of two-point
correlation function). Another peculiarity related to this region is the following.
At fixed $\kp$, $0<\kp\le 1$, in the region where $h>(1-\kp^2)^{1/2}$ the NS-vacuum energy is lower than
R-vacuum energy (asymptotically, at $n\to\infty$, they become coinciding).
In the region $0\le h\le (1-\kp^2)^{1/2}$ there are intersections at special values of $h$ of these vacuum levels even at finite $n$.
The number of these intersections grows with $n$.
For a detailed analysis of the oscillatory region see \cite{BarouchMcCoy,HGR}.

In this paper we derive the formulas for the matrix elements in paramagnetic phase ($h>1$, $0<\kp<1$)
and add comments on the modification of the formulas for other values of parameters.

\subsection{Eigenvalues and eigenvectors of the Hamiltonian of XY-chain}
\label{cleval}

Using Jordan--Wigner and Bogoliubov transformations
the Hamiltonian ${\cal H}$ of the XY-chain
can be rewritten as the Hamiltonian of the system of free fermions and diagonalized \cite{LSM,Katsura}.
The relation between energies $\ve(\qu)$ and momenta $q$ of the fermionic excitations is
\be\label{enXY}
\ve(\qu)=\left((h-\cos\qu)^2+\kp^2\sin^2\qu\right)^{1/2},\qquad \qu\ne 0,\pi\,,
\ee
\[\ve(0)=h-1,\quad\quad \ve(\pi)=h+1\,.\]

The Hamiltonian ${\cal H}$ commutes with the operator ${\bf V}= \sz{1} \sz{2} \cdots \sz{n}$.
Since  ${\bf V}^2=1$, the eigenvectors are separated to two sectors with respect to the eigenvalue of ${\bf V}$.
Below the sign $+/-$ in front of $\ve(q)$ in the expression for energies ${\cal E}$ corresponds
to the absence/presence of the fermionic excitation
with the momentum $q$. Each such excitation carries the energy $\ve(q)$.
\begin{itemize}
\item
NS--sector:  ${\bf V}\to +1 $, the fermionic excitations have ``half-integer'' quasimomenta
\[
q\in {\rm NS}=\left\{\frac{2\pi}{n}(j+1/2)\,, \ j\in \mathbb{Z}_n\right\}\quad \Rightarrow \quad
{\cal E}=-\frac{1}{2}\sum_{q\in {\rm NS}}\pm {\varepsilon(q)}\,.
\]
This sector includes the states only with an even number of excitations.

\item
R--sector:    ${\bf V}\to -1 $, the fermionic excitations have ``integer'' quasimomenta
\be
q\in {\rm R}=\left\{\frac{2\pi}{n}j\,, \ j\in \mathbb{Z}_n\right\}\quad \Rightarrow \quad
{\cal E}=-\frac{1}{2}\sum_{q\in {\rm R}}\pm {\varepsilon(q)}\,. \label{energyR}
\ee
In the paramagnetic phase this sector includes the states only with an odd number of excitations.
In the ferromagentic phase ($0\le h<1$) it is natural to re-define the
energy of zero-momentum excitation as $\ve(0)=1-h$ to be positive. {}From the formula \rr{energyR} for energy ${\cal E}$, this change
of the sign of $\ve(0)$ in the ferromagnetic phase leads to a formal change between
absence/presence of zero-momentum excitation in the labelling of eigenstates.
Thus although the analytical expressions for energies ${\cal E}$ in terms of $h$ and $\kp$ are the same in both phases,
because of the redefinition of $\ve(0)$ in the case of $0\le h<1$ the number of excitations in the ferromagnetic phase is even.

\end{itemize}

We will denote the eigenstates $|\Phi\rangle^\alpha_{\rm model}$ by the set of values of the excited quasi-momenta $\Phi=\{q_1,q_2,\ldots,q_L\}$,
the label of the sector $\alpha=$ NS or R and the label of the model.
For example, the state in R-sector of the quantum XY-chain with $n=3$ sites in the paramagnetic phase
with all possible quasi-momenta $\Phi=\{0,2\pi/3,-2\pi/3\}$ excited is $|0,2\pi/3,-2\pi/3\rangle_{XY}^{\rm R}$.
It has the energy
\[{\cal E}=-\frac 1 2 \left(1-h-\ve(2\pi/3)-\ve(-2\pi/3)\right)\,.
\]
The same formula for the energy in the ferromagnetic phase corresponds to the state $|2\pi/3,-2\pi/3\rangle_{XY}^{\rm R}$.

\subsection{Matrix elements of spin operators}

Formally in order to calculate any correlation function for XY-chain it is sufficient to find the matrix elements of
spin operators $\sx{k}$, $\sy{k}$ and $\sz{k}$ between the eigenstates of Hamiltonian $\cal H$.

\begin{itemize}
\item {\em Matrix elements of $\sigma^{\rm z}_k$.}

The operator $\sigma^{\rm z}_k$ commutes with ${\bf V}= \sz{1} \sz{2} \cdots \sz{n}$. Therefore the action of $\sigma^{\rm z}_k$
does not change the sector. In fact the operator $\sigma^{\rm z}_k$ can be presented as a bilinear combination of
operators of creation and annihilation of the fermionic excitations. Hence the matrix elements of $\sigma^{\rm z}_k$
between eigenvectors of ${\cal H}$ can be calculated easily (most of them are $0$).
We will not consider such matrix elements in this paper.

\item {\em Matrix elements of $\sigma^{\rm x}_k$ and $\sigma^{\rm y}_k$.}

The operators $\sigma^{\rm x}_k$ and $\sigma^{\rm y}_k$ anticommute with
${\bf V}= \sz{1} \sz{2} \cdots \sz{n}$. Therefore their action changes the sector.
The operators $\sigma^{\rm x}_k$ and $\sigma^{\rm y}_k$
can not be presented in terms of fermionic operators in a local way. All the matrix elements of them
between the eigenvectors of ${\cal H}$ from different sectors are non-zero!

\end{itemize}

The aim of this paper is to derive explicit factorized formula for the
matrix elements of $\sigma^{\rm x}_k$ and $\sigma^{\rm y}_k$.
The idea is to relate three models: the quantum XY-chain in a transverse field, the Ising model on 2D lattice and
$N=2$ BBS model. The relation between the first and the second model is based on the observation by
M.~Suzuki \cite{Suzuki}. The relation between the second and the third is based on \cite{BIS}. The latter relation together
with the results on separation of variables for BBS model allowed \cite{gipst2} to prove
the formulas for the matrix elements of spin operator of Ising model found by A.~Bugrij and O.~Lisovyy \cite{BL1,BL2}.
In this paper we transfer these results on the matrix elements to the case of XY-chain.
The parameters of these three models are ($h,\kp$), ($K_x,K_y$) and ($a,b$), respectively.

\section{Relation between three models}

\subsection{Relation between quantum XY-chain and the Ising model on a lattice}

The row-to-row transfer-matrix of the two-dimensional Ising model with parameters $K_x$ and $K_y$
 can be chosen as
\be\label{tIs}
\fl {\bf t}_{\rm XY}:=T_1^{1/2}T_2T_1^{1/2}=
\exp{\lk {\sum_{k=1}^n}\, \frac{K^*_y}{2}\, \sz{k}\rk}\:
\exp{ \lk{\sum_{k=1}^n}\, K_x \,\sx{k}\,\sx{k+1}\rk}\:
\exp{\lk {\sum_{k=1}^n}\, \frac{K^*_y}{2}\, \sz{k}\rk}\,,
\ee
where the spin configurations of the rows are chosen to be labeled by the eigenvectors of the operators $\sx{k}$,
the parameter $K_y^*$ is dual to $K_y$, that is $\tanh K_y=\exp(-2K_y^*)$,  and
\be\label{T1T2}
T_1=\exp{\lk {\sum_{k=1}^n}\, K^*_y\, \sz{k}\rk}\:,
\qquad
T_2=\exp{ \lk{\sum_{k=1}^n}\, K_x \,\sx{k}\,\sx{k+1}\rk}\:.
\ee

In \cite{Suzuki}, M.~Suzuki observed that if we choose $K_x$ and $K^*_y$ such that
\be\label{relXY-Is}
\tanh 2K_x=\frac{\sqrt{1-\kp^2}}{h}\,,\qquad
\cosh 2K^*_y=\frac{1}{\kp}\,
\ee
then the Hamiltonian \rr{HamXYh} of XY-chain will commute with the transfer-matrix of the 2D Ising model \rr{tIs} and these two
operators have a common set of eigenvectors.

\subsection{$N=2$ BBS model and its relation to the Ising model}

To define $N=2$ BBS model we use the following $L$-operator\footnote{\label{sxsz}In comparison
with \cite{gipst2} we interchanged $\sx{k}$ and $\sz{k}$ in the $L$-operator
\rr{L-Ising}. It is just another representation of the Weyl algebra entering the definition of $L$-operator.}  \cite{BS,Kore}
\be\label{L-Ising}
L_k(\lm)=\lk\ba{cc} 1\,+\lm\,\sz{k} &  \lm \,\sx{k}\, (a\,-b\, \sz{k})\\[2mm]
\sx{k}\, (a\,-b\, \sz{k}) & \lm a^2 \,+ \sz{k}\, b^2\ea \rk\,,\ee
depending on parameters $a$, $b$ and spectral parameter $\lm$.
It satisfies the Yang--Baxter equation with (twisted) quantum trigonometric $R$-matrix.
In particular it means that the eigenvectors of the transfer matrix
${\bf t}(\lm)={\rm tr}\, L_1(\lm)\,L_2(\lm)\cdots L_n(\lm)$ built from such $L$-operators are
independent of $\lm$.

Fixing the spectral parameter to the value $\lm=b/a$, the $L$-operator \rr{L-Ising} degenerates
\[  L_k(b/a)\;=\;(1\,+\, \sz{k}\:b/a)\lk \ba{c} 1\\ a\, \sx{k} \ea \rk\lk \ba{cc} 1\,,& b\,\sx{k}\ea \rk \]
and the transfer matrix ${\bf t}(\lm)$ can be put into a nonsymmetric Ising form
\bea\fl
{\bf t}(b/a)=
\prod_{k=1}^n (1+\sz{k} \cdot {b}/{a})\cdot \prod_{k=1}^n (1+\sx{k}\sx{k+1}\cdot a\,b)
=(\cosh K_x\cosh K^*_y)^{-n}\, T_1T_2\,,\ny\\\fl
\label{Isitra}
 {\bf t}(b/a)  \sim\:T_1 T_2=\exp{\lk {\textstyle\sum_{k=1}^n}\, K^*_y\, \sz{k}\rk}\:
\exp{ \lk{\textstyle\sum_{k=1}^n}\, K_x \,\sx{k}\,\sx{k+1}\rk}\,,
\eea
if we use periodic boundary condition $\sigma^i_{n+k}=\sigma^i_{k}$ and identify
\be
e^{-2K_y}\,= \tanh K^*_y\,=\,{b}/{a}\,,\quad\; \tanh K_x\,=\,a\!\,b\,.
\label{relIs-BBS}\ee
Thus at $\lm=b/a$ we get the transfer-matrix of the Ising model. If we do not fix
the spectral parameter to this special value, we shall talk of the ``generalized Ising model''. However,
transfer matrix eigenstates are independent of the choice of $\lm$.
In \cite{gips,gipst1,gipst2} the eigenvectors for
the nonsymmetric transfer matrix \rr{Isitra} and the matrix elements of $\sx{k}$ between these eigenvectors were derived using
method of separation of varibales.

Comparing \rr{relXY-Is} and \rr{relIs-BBS} we get the following simple relations for the parameters of XY-model
and special BBS-model
with $L$-operator \rr{L-Ising}:
\be\label{relXY-BBS}
\kp= \frac{a^2 - b^2}{a^2 + b^2}\,,\qquad
h= \frac{1 + a^2 b^2}{a^2 + b^2}\,.
\ee

\subsection{Relation between the energies of excitations for the Ising model and XY-chain}

In the previous subsections we have shown how the quantum XY-chain in a transverse field, the Ising model on 2D lattice and
$N=2$ BBS model are related.
The parameters of the models are ($h,\kp$), ($K_x,K_y$) and ($a,b$), respectively.
The relations  between these pairs of parameters are given by \rr{relXY-Is}, \rr{relIs-BBS} and \rr{relXY-BBS}.

Since the transfer-matrices
${\bf t}_{\rm XY}=T_1^{1/2}T_2T_1^{1/2}$, ${\bf t}_{\rm Is}=T_2^{1/2}T_1 T_2^{1/2}$, $T_1T_2$
of 2D Ising model are related by similarity transformations,
the enumeration of the eigenstates of all these transfer-matrices is the same as described in Sect.~2.2 for the quantum XY-chain.
They will be denoted, respectively, by $|\Phi\rangle^\alpha_{\rm XY}$, $|\Phi\rangle^\alpha_{\rm Is}$, $|\Phi\rangle^\alpha$,
where $\Phi=\{p_1,p_2,\ldots,p_L\}$ is the set of values of the excited quasi-momenta $\Phi=\{p_1,p_2,\ldots,p_L\}$
and $\alpha=$ NS or R  is the label of the sector.
Their eigenvalues $e^{-\gamma_\Phi}$ are the same: 
\be\label{eval}\fl
{\bf t}_{\rm XY} |\Phi\rangle^\alpha_{\rm XY}=e^{-\gamma_\Phi}|\Phi\rangle^\alpha_{\rm XY}\,,\quad
{\bf t}_{\rm Is}|\Phi\rangle^\alpha_{\rm Is}=e^{-\gamma_\Phi}|\Phi\rangle^\alpha_{\rm Is}\,,\quad
T_1T_2|\Phi\rangle^\alpha=e^{-\gamma_\Phi}|\Phi\rangle^\alpha\,,
\ee\be\label{gamPhi}
\gamma_\Phi=\sum_{l=1}^L\gamma(p_l)-\frac{1}{2}\sum_{p\in \alpha} \gamma(p)\,,
\ee\be\label{drIs}
\cosh\g(p) = \frac{(t_x + t_x^{-1})(t_y + t_y^{-1})}{2(t_x^{-1} - t_x)} -
\frac{t_y^{-1}-t_y}{t_x^{-1}-t_x} \cos p\,,
\ee\[
t_x=\tanh K_x\,, \qquad t_y=\tanh K_y\,.
\]

The eigenvalues of the transfer matrix ${\bf t}(\lm)$ of the BBS model with $L$-operator \rr{L-Ising}
are proportional to $\prod_p (\lm\pm s_p)$ (see formula (68) of \cite{gipst1}),
\be\label{drBBS}
s_p=\left(\frac{b^4-2b^2\cos p+1}{a^4-2a^2\cos p+1}\right)^{1/2}\,,
\ee
where the sign $+/-$ in the front of $s_p$ in the expression for the eigenvalues of ${\bf t}(\lm)$ corresponds
to the absence/presence of the fermionic excitation with the momentum $p$.
The momentum $p$ runs over the same sets as in the case of the quantum XY-chain. Due to \rr{Isitra}, \rr{eval} and \rr{gamPhi}
we have the relation  between $\g(p)$ and $s_p$ (see \cite{gipst2}):
\be\label{erIsBBS}
e^{\g(p)}=\frac{a s_p+b}{a s_p-b}
\ee
and a relation between $\ve(p)$ and $\g(p)$:
using \rr{relXY-BBS}, \rr{drBBS} and \rr{erIsBBS} we get
\[\fl
\sinh \g(p)= \frac{2ab s_p}{a^2 s_p^2-b^2}=\frac{2ab}{(a^2-b^2)(1-a^2b^2)}
\sqrt{(b^4-2b^2\cos p+1)(a^4-2a^2\cos p+1)}
\]\be
=\frac{2ab(a^2+b^2)}{(a^2-b^2)(1-a^2b^2)}\, \ve(p)=
\frac{\sqrt{1-\kp^2}}{\kp\sqrt{\kp^2+h^2-1}}\, \ve(p)\,. \label{relge}
\ee
The existence of the relation between $\g(p)$ and $\ve(p)$ is surprising because the commu\-ta\-ti\-vity of
the Hamiltonian \rr{HamXYh} of the XY-chain and the transfer matrix \rr{tIs} of the 2D Ising model
does not imply {\em a priori} any relation between their eigenvalues.

\subsection{Uniformization of the dispersion relation \rr{drIs}}
\label{elparam}

We use a parametrization of the dispersion relation \rr{drIs} of the 2D Ising model
in terms of elliptic function at $h>1$, $0<\kp<1$ which corresponds to the paramagnetic phase of Ising model. This parametrization is a
modification of parametrization from \cite{Palmer} given for the ferromagnetic phase of Ising model
and corresponding to $0<h<1$, $(1-h^2)^{1/2}<\kp<1$ for XY-chain.

We introduce the modulus of elliptic curve ${\sf k}^{-1}$ by
\[
{\sf k}^{-1}=\sinh 2K_x \sinh 2 K_y=\kp/\sqrt{\kp^2+h^2-1}=(a^2-b^2)/(1-a^2b^2)\,.
\]
In the paramagnetic phase we have $0\le {\sf k}^{-1} <1$.
Complete elliptic integrals for ${\sf k}^{-1}$ and for the supplementary modulus are
$K=K({\sf k}^{-1})$ and $K'=K((1-{\sf k}^{-2})^{1/2})$, respectively.
We define  real parameter ${\sf a}$, $0<{\sf a}<K'/2$, by one of the equivalent relations
\[
1/\sinh 2K_x={\rm i} {\sf k} / \mbox{sn} (2 {\rm i} {\sf a}, {\sf k}^{-1}),\qquad
1/\sinh 2K_y=-{\rm i}\;  \mbox{sn} (2 {\rm i} {\sf a}, {\sf k}^{-1})\,.
\]
The following two elliptic functions
\[\fl
\lm(u)= \mbox{sn} (u- {\rm i} {\sf a}, {\sf k}^{-1})/ \mbox{sn} (u+ {\rm i} {\sf a}, {\sf k}^{-1})\,,\quad
z(u)= {\sf k}^{-1}\;\mbox{sn} (u- {\rm i} {\sf a}, {\sf k}^{-1})\; \mbox{sn} (u+ {\rm i} {\sf a}, {\sf k}^{-1})
\]
satisfy the relation
\be\label{lz}
\sinh 2K_x\, (z + z^{-1}) + \sinh 2K_y\, (\lm + \lm^{-1}) =  2 \cosh 2K_x \cosh 2K_y \,.
\ee
To prove it we note that the left-hand side of \rr{lz} is an elliptic function without poles
(the poles at $\pm{\sf a}$ and $\pm{\sf a}+{\rm i} K'$ are canceled) and therefore it is a constant. Thus it is sufficient
to establish validity of \rr{lz} at $u=0$. For this end we use $\lm(0)=-1$,
\[\fl
z(0)=-{\sf k}^{-1}\;\mbox{sn}^2 ( {\rm i} {\sf a}, {\sf k}^{-1})=
-{\sf k}\;\frac{1-\mbox{dn} ( 2 {\rm i} {\sf a}, {\sf k}^{-1})}{1+\mbox{cn} ( 2 {\rm i} {\sf a}, {\sf k}^{-1})}\,,
\qquad
z^{-1}(0)=-{\sf k}\;\frac{1+\mbox{dn} ( 2 {\rm i} {\sf a}, {\sf k}^{-1})}{1-\mbox{cn} ( 2 {\rm i} {\sf a}, {\sf k}^{-1})}
\]
which follow from the formulas of Example~6, Sect.~22.21 of \cite{WhitWat}, and
\[
\cosh 2K_x = \mbox{dn} ( 2 {\rm i} {\sf a}, {\sf k}^{-1})\, , \qquad
\cosh 2K_y = {\rm i}\; \mbox{cn} ( 2 {\rm i} {\sf a}, {\sf k}^{-1})/\mbox{sn} ( 2 {\rm i} {\sf a}, {\sf k}^{-1})\,.
\]

The relation \rr{lz} coincides with the dispersion relation \rr{drIs} if one identifies $z(u)=e^{-{\rm i} p}$ and $\lm(u)=e^{-\g(p)}$.
The parameter $u$ on the elliptic curve is an analogue of rapidity.
Now if $p$ runs from $-\pi$ to $\pi$ then $u$ runs along the segment from ${\rm i}K'/2$ to $2K+{\rm i}K'/2$.

There is another dispersion relation corresponding to the evolution in the transverse direction on the Ising lattice:
\be\label{drIsdual}
\cosh\bar\g(\bar p) = \frac{(t_x + t_x^{-1})(t_y + t_y^{-1})}{2(t_y^{-1} - t_y)} -
\frac{t_x^{-1}-t_x}{t_y^{-1}-t_y} \cos \bar p\,.
\ee
It is uniformized by $\lm(u)=e^{{\rm i} \bar p}$ and $z(u)=e^{-\bar\g(\bar p)}$.
Now if $\bar p$ runs from $-\pi$ to $\pi$ then $u$ runs along the segment from $0$ to $2K$.


{}From \rr{relXY-BBS}, we have
\be\label{abhkp}
a^2=\frac{h-\sqrt{h^2+\kp^2-1}}{1-\kp}\,,\qquad b^2=\frac{h-\sqrt{h^2+\kp^2-1}}{1+\kp}\,.
\ee
The points with $z=a^{\pm 2}$ and $z=b^{\pm 2}$ are the branching points of the spectral curve \rr{lz} considered as $\lm(z)$.
The parameters $a^2$, $b^2$ correspond respectively to $\lm_2$, $\lm_1^{-1}$ of \cite{BarouchMcCoy} and
to $\alpha^{-1}_1$, $\alpha^{-1}_2$ of \cite{Palmer}.
We have also $a^2=e^{-\bar\g(0)}$, $b^2=e^{-\bar\g(\pi)}$.

\section{Formulas for the matrix elements of spin operators}
\label{MEsection}

In this section we will derive the formulas for the matrix elements of spin operators for quantum XY-chain of finite length.
The derivation for the basic region of parameters $h>1$, $0<\kp<1$ is given in Sect.~\ref{sectpar}.
The formulas for other values of parameters can be obtained by
analytic continuation. The details of the continuation are given in the following subsections.

\subsection{Paramagnetic phase: $h>1$, $0<\kp<1$}
\label{sectpar}

We use the Bugrij--Lisovyy formula ((40) of \cite{BL2}) for the matrix element of spin operator between the eigenvectors
$|\Phi_0\ra_{\rm Is}=|\qu_1, \qu_2,\ldots,\qu_K\ra_{\rm Is}^{\rm NS}$
and $|\Phi_1\ra_{\rm Is}=|\pu_1, \pu_2,\ldots,\pu_L\ra_{\rm Is}^{\rm R}$ of the symmetric transfer matrix ${\bf t}_{\rm Is}=T_2^{1/2}T_1 T_2^{1/2}$
for the finite 2D Ising model (the states are labeled by the momenta of excited fermions as it is explained in Sects.~\ref{cleval} and~3.3):
\[\fl
\Xi_{\Phi_0,\Phi_1}=|\,{}^{\rm NS}_{\rm Is}\langle\, \qu_1, \qu_2,\ldots,\qu_K\,|\;\sx{m}\;|\,\pu_1,\pu_2,\ldots,\pu_L\,\rangle_{\rm Is}^{\rm R}|^2
\]\[\fl
=\;\xi\; \xi_T\; \prod_{k=1}^K \;\frac{\prod^{\rm NS}_{\qu\ne \qu_k} \sinh \frac{\g(\qu_k)+\g(\qu)}{2}}
{n \prod^{\rm R}_{\pu} \sinh \frac{\g(\qu_k)+\g(\pu)}{2}}\;\;
\prod_{l=1}^L \;\frac{\prod^{\rm R}_{\pu\ne \pu_l} \sinh \frac{\g(\pu_l)+\g(\pu)}{2}}
{n \prod^{\rm NS}_{\qu} \sinh \frac{\g(\pu_l)+\g(\qu)}{2}}\cdot
\left(\frac{t_y-t_y^{-1}}{t_x-t_x^{-1}}\right)^{\!\!(K-L)^2/2}
\]\be\fl\qquad \times\;
\prod_{k<k'}^K \frac{\sin^2\frac{\qu_k-\qu_{k'}}{2}} {\sinh^2 \frac{\g(\qu_k)+\g(\qu_{k'})}{2}}
\;\;\prod_{l<l'}^L \frac{\sin^2\frac{\pu_l-\pu_{l'}}{2}} {\sinh^2 \frac{\g(\pu_l)+\g(\pu_{l'})}{2}}
\prod_{1\le k \le K \atop 1\le l \le L}
\frac {\sinh^2 \frac{\g(\qu_k)+\g(\pu_l)}{2}} {\sin^2\frac{\qu_k-\pu_l}{2}}\,, \label{ME_BL}
\ee \be\label{xi}
\fl \xi^4={\sf k}^2-1=(\sinh 2K_x \sinh 2K_y)^{-2}-1=\frac{h^2-1}{\kp^2}=\frac{(1-a^4)(1-b^4)}{(a^2-b^2)^2}\,,
\ee\be\label{xiT}
\xi_T^4=
\frac{\prod^{\rm NS}_{\qu} \prod^{\rm R}_{\pu} \sinh^2 \frac{\g(\qu)+\g(\pu)}{2}}
{\prod^{\rm NS}_{\qu,\qu'} \sinh \frac{\g(\qu)+\g(\qu')}{2}
\prod^{\rm R}_{\pu,\pu'} \sinh \frac{\g(\pu)+\g(\pu')}{2}}\,,
\ee\[
t_x=\tanh K_x=ab\,,\qquad t_y=\tanh K_y=\frac{a-b}{a+b}\,,\]
\be\label{tytx}
\frac{t_y-t_y^{-1}}{t_x-t_x^{-1}}=\frac{4a^2b^2}{(a^2-b^2)(1-a^2 b^2)}=\frac{1-\kp^2}{\kp\sqrt{\kp^2+h^2-1}}\,,
\ee
where we used \rr{relXY-Is}, \rr{relIs-BBS} and \rr{relXY-BBS} to write equivalent expressions in terms of different parameters.

Note that the Bugrij--Lisovyy formula \rr{ME_BL} is given for the {\em normalized} eigenvectors $|\Phi\ra_{\rm Is}$ of the transfer matrix
${\bf t}_{\rm Is}=T_2^{1/2}T_1 T_2^{1/2}$ which differs from the transfer matrix \rr{tIs}.
The eigenvectors $|\Phi\ra_{\rm XY}$ of the Hamiltonian \rr{HamXYh} of XY-chain and \rr{tIs} are in one-to-one correspondence
with the eigenvectors $|\Phi\ra_{\rm Is}$ and the eigenvectors $|\Phi\ra$ of BBS model. All the eigenvectors
with the same $\Phi$ (the same set of excited fermion excitations) are related by  similarity transformations.

In \cite{gips,gipst1} the left and right eigenvectors $\la \Phi|$ and $|\Phi\ra$ of $T_1T_2$ were found.
They are related to $|\Phi\ra_{\rm Is}$ by the action of operator $T_2^{1/2}$ and its inverse. Since these operators commute with $\sx{m}$,
we have the relation
\be\label{relMEIsBBS}
\Xi_{\Phi_0,\Phi_1}=\frac{\la \Phi_0|\sx{m}|\Phi_1\ra \la \Phi_1|\sx{m}|\Phi_0\ra}{\la \Phi_0|\Phi_0\ra \la \Phi_1|\Phi_1\ra}
\ee
expressing the matrix elements of 2D Ising model in terms of the matrix elements of BBS model found in \cite{gipst2}
and used to prove \rr{ME_BL}.

Here we want to use the matrix elements of BBS model to derive the matrix elements of spin operators
between the eigenstates of XY quantum chain.
Since the Hamiltonian of XY-chain and the transfer matrix ${\bf t}_{\rm Is}$ are Hermitian matrices, there is a natural
way  to relate the left and right eigenvectors and to normalize them.
In the case of nonsymmetric transfer matrix $T_1 T_2$ arising in BBS model,
the left and right eigenvectors $\la \Phi|$ and $|\Phi\ra$ are unrelated
but we will relate them to normalized eigenvectors of XY-chain by
$\alpha^{\rm L}_\Phi\cdot {}_{\rm XY}\la \Phi|=\la \Phi|T_1^{1/2}$ and  $\alpha^{\rm R}_\Phi\cdot |\Phi\ra_{\rm XY}=T_1^{-1/2}|\Phi\ra$.
We fix $\alpha^{\rm R}_\Phi=||T_1^{-1/2}|\Phi\ra||>0$, then the coefficient $\alpha^{\rm L}_\Phi$ is determined
from the requirement ${}_{\rm XY}\la \Phi|=|\Phi\ra_{\rm XY}^\dag$.
We have also $\alpha^{\rm L}_{\Phi} \alpha^{\rm R}_{\Phi}=\la \Phi|\Phi\ra$.
Since $\la \Phi|\Phi\ra$ is real at real $a$ and $b$, both $\alpha^{\rm L}_{\Phi}$ and $\alpha^{\rm R}_{\Phi}$ are real too.
The matrix elements of BBS model are related to the matrix elements of XY-chain by
\be\label{relME1}
\fl \la \Phi_0|\sx{m}|\Phi_1\ra=\alpha^{\rm L}_{\Phi_0} \alpha^{\rm R}_{\Phi_1} \cdot
{}_{\rm XY}\la \Phi_0|T_1^{-1/2}\sx{m}T_1^{1/2}|\Phi_1\ra_{\rm XY}
\ee\be\label{relME2}
=\alpha^{\rm L}_{\Phi_0} \alpha^{\rm R}_{\Phi_1} \cdot e^{\gamma_{\Phi_0}-\gamma_{\Phi_1}}
 {}_{\rm XY}\la \Phi_0|T_1^{1/2}\sx{m}T_1^{-1/2}|\Phi_1\ra_{\rm XY}\,,\ee
where the last relation follows from the facts that $|\Phi\ra_{\rm XY}$ is eigenvector of
$T_1^{1/2}T_2T_1^{1/2}$ with the eigenvalue $e^{-\gamma_\Phi}$ (see  \rr{eval}) and $T_2$ commutes with $\sx{m}$.
Complex conjugation of \rr{relME2} together with \rr{relME1} with interchanged $\Phi_0$ and $\Phi_1$ give
\be\label{relME3}
\overline{\la \Phi_0|\sx{m}|\Phi_1\ra}=\frac{\alpha^{\rm L}_{\Phi_0} \alpha^{\rm R}_{\Phi_1}}{\alpha^{\rm L}_{\Phi_1} \alpha^{\rm R}_{\Phi_0}}
\cdot e^{\gamma_{\Phi_0}-\gamma_{\Phi_1}}
 {}_{\rm XY}\la \Phi_0|T_1^{1/2}\sx{m}T_1^{-1/2}|\Phi_1\ra_{\rm XY}\,.
\ee

{}From the other side, the formulas (56) and (57) of \cite{gipst2} give the following factorized presentations for the matrix element
between eigenstates of the transfer matrix $T_1 T_2$  (see  footnote on p.~\pageref{sxsz}):
\[
\la \Phi_0|\sx{m}|\Phi_1\ra=f_1(b) f_2(b^2)\,,\qquad
\la \Phi_1|\sx{m}|\Phi_0\ra=f_1(-b) \overline{f_2(b^2)}\,,
\]
where $f_1(b)$ is a real (we suppose that $a$ and $b$ are real) and $f_2(b^2)$ is a complex
but invariant with respect to $b\to -b$. Therefore
\be\label{ratioME}
\frac{\ \la \Phi_0|\sx{m}|\Phi_1\ra\ }{\overline{\la \Phi_1|\sx{m}|\Phi_0\ra}}=\frac{f_1(b)}{f_1(-b)}=:C_{\Phi_0,\Phi_1}\,.
\ee
It is easy to prove using explicit formulas for $f_1(b)$ from \cite{gipst2} that
\be\label{C}
C_{\Phi_0,\Phi_1}=\frac{\prod_{\pu\in {\rm R}} e^{\g(\pu)/2}}
{\prod_{\qu\in {\rm NS}} e^{\g(\qu)/2}} \frac{\prod_{k=1}^K e^{\g(\qu_k)}}{\prod_{l=1}^L e^{\g(\pu_l)}}=e^{\gamma_{\Phi_0}-\gamma_{\Phi_1}}
\ee
for $|\Phi_0\ra=|\qu_1, \qu_2,\ldots,\qu_K\ra^{\rm NS}$
and $|\Phi_1\ra=|\pu_1, \pu_2,\ldots,\pu_L\ra^{\rm R}$.
Let us consider, for example, the case of odd $n$ (the length of the chain) and $\sigma_0=\sigma_\pi$
($\sigma_q=0$/$\sigma_q=1$ corresponds to absence/presence of fermion excitation with momentum $q$).
{}From Eq.~(56) and the discussion in Section~6.1 of \cite{gipst2} we can choose
\[
f_1(b)=((-1)^{\sigma_0}-a\,b)
\prod_{k\in \check{\cal D}} ((-1)^k b+a s_{\pi k/n}) \prod_{k\in \hat{\cal D}} ((-1)^k b-a s_{\pi k/n})\,,
\]
where $s_p$ is given by \rr{drBBS} and the set $\hat{\cal D}$ (resp. $\check{\cal D}$) consists of such $k$ from
$\{1,2,\ldots,n-1\}$ for which the fermions with both momenta $\pm\pi k/n$  are excited (resp. not excited)
in the states $|\Phi_0\ra$ and $|\Phi_1\ra$. Using \rr{erIsBBS} we get
\[
C_{\Phi_0,\Phi_1}=\frac{f_1(b)}{f_1(-b)}=\frac{(-1)^{\sigma_0}-a\,b}{(-1)^{\sigma_0}+a\,b}\,\cdot
\frac{\prod_{\pu\in {\rm R},\pu\ne 0} e^{\g(\pu)/2}} {\prod_{\qu\in {\rm NS}, \qu\ne\pi} e^{\g(\qu)/2}}\cdot
\frac{\prod_{k=1, \qu_k\ne \pi}^K e^{\g(\qu_k)}} {\prod_{l=1, \pu_l\ne 0 }^L e^{\g(\pu_l)}}\,.
\]
Thus it remains to verify that the first fraction also fits \rr{C} as contribution of the momenta $0$ and $\pi$.
For this end we need just to take into account \rr{drBBS}, \rr{erIsBBS} and
\[
s_0=\frac{b^2-1}{a^2-1}\,, \qquad
s_\pi=\frac{b^2+1}{a^2+1}\,,\qquad
e^{(\g(0)-\g(\pi))/2}=\frac{1-a\,b}{1+a\,b}\,.
\]
The correct sign of the latter formula can be fixed from the limit $b=0$ (the quantum Ising chain limit) and then
taking limit $a\to 0$ (it corresponds to the limit of strong external field $h\to \infty$).
All the other three cases of odd (even) $n$ and
$\sigma_0=\sigma_\pi$ ($\sigma_0\ne \sigma_\pi$) can be analyzed similarly. It proves \rr{C}.

Taking into account \rr{relMEIsBBS} and  \rr{ratioME}  we get
\be\label{MEaf}
\frac{\la \Phi_0|\sx{m}|\Phi_1\ra}{\bigl(\la \Phi_0|\Phi_0\ra \la \Phi_1|\Phi_1\ra\bigr)^{1/2}}
=e^{{\rm i}\delta_{\Phi_0,\Phi_1}}\left(C_{\Phi_0,\Phi_1}\,\Xi_{\Phi_0,\Phi_1}\right)^{1/2}\,,
\ee
where $\delta_{\Phi_0,\Phi_1}$ is a phase related to a particular normalization of eigenvectors.
To relate these matrix elements to the matrix elements of XY-chain we observe that
\rr{relME3} and \rr{ratioME} imply
$\alpha^{\rm L}_{\Phi_0} \alpha^{\rm R}_{\Phi_1}=\alpha^{\rm L}_{\Phi_1} \alpha^{\rm R}_{\Phi_0}$.
Since $\alpha^{\rm L}_{\Phi_0} \alpha^{\rm R}_{\Phi_0}\alpha^{\rm L}_{\Phi_1}\alpha^{\rm R}_{\Phi_1} = \la \Phi_0|\Phi_0\ra\la \Phi_1|\Phi_1\ra$
we obtain $\alpha^{\rm L}_{\Phi_0} \alpha^{\rm R}_{\Phi_1}=\bigl(\la \Phi_0|\Phi_0\ra\la \Phi_1|\Phi_1\ra\bigr)^{1/2}$.
Thus from \rr{relME1}, \rr{relME2}, \rr{C} and \rr{MEaf} we derive
\[\fl
{}_{\rm XY}\la \Phi_0|T_1^{-1/2}\sx{m}T_1^{1/2}|\Phi_1\ra_{\rm XY}
\]\[={}_{\rm XY}\la \Phi_0|\lk \sx{m} \cosh K^*_y -{\rm i} \sy{m} \sinh K^*_y \rk|\Phi_1\ra_{\rm XY}=
e^{{\rm i}\delta_{\Phi_0,\Phi_1}}\left(C_{\Phi_0,\Phi_1}\,\Xi_{\Phi_0,\Phi_1}\right)^{1/2}\,,
\]\[\fl
{}_{\rm XY}\la \Phi_0|T_1^{1/2}\sx{m}T_1^{-1/2}|\Phi_1\ra_{\rm XY}
\]\[
={}_{\rm XY}\la \Phi_0|\lk \sx{m} \cosh K^*_y +{\rm i} \sy{m} \sinh K^*_y \rk|\Phi_1\ra_{\rm XY}=
e^{{\rm i}\delta_{\Phi_0,\Phi_1}}\left(C^{-1}_{\Phi_0,\Phi_1}\,\Xi_{\Phi_0,\Phi_1}\right)^{1/2}\,.
\]
Finally taking appropriate linear combinations of these two formulas we get
the main result of the paper:
the matrix elements of spin operators between the eigenvectors
$|\Phi_0\ra_{\rm XY}=|\qu_1, \qu_2,\ldots,\qu_K\ra_{\rm XY}^{\rm NS}$ from the NS-sector
and $|\Phi_1\ra_{\rm XY}=|\pu_1, \pu_2,\ldots,\pu_L\ra_{\rm XY}^{\rm R}$ from the R-sector of the Hamiltonian \rr{HamXYh} of XY-chain are
\[
|{}_{\rm XY}\la \Phi_0|\sx{m}|\Phi_1\ra_{\rm XY}|^2=
\frac{\kp}{2(1+\kp)} \left(C_{\Phi_0,\Phi_1}^{1/2}+ C_{\Phi_0,\Phi_1}^{-1/2}\,\right)^2 \Xi_{\Phi_0,\Phi_1}
\]\be\label{MEsx}
\qquad=\frac{2\kp}{1+\kp} \cosh^2\frac{\gamma_{\Phi_0}-\gamma_{\Phi_1}}{2}\, \Xi_{\Phi_0,\Phi_1}\,,
\ee\[
|{}_{\rm XY}\la \Phi_0|\sy{m}|\Phi_1\ra_{\rm XY}|^2=\frac{\kp}{2(1-\kp)}
\left(C_{\Phi_0,\Phi_1}^{1/2}- C_{\Phi_0,\Phi_1}^{-1/2}\,\right)^2 \Xi_{\Phi_0,\Phi_1}
\]\be\label{MEsy}
\qquad=\frac{2\kp}{1-\kp} \sinh^2\frac{\gamma_{\Phi_0}-\gamma_{\Phi_1}}{2}\, \Xi_{\Phi_0,\Phi_1}\,,
\ee
where $\Xi_{\Phi_0,\Phi_1}$, $C_{\Phi_0,\Phi_1}$ and $\gamma_\Phi$
are given by \rr{ME_BL}, \rr{C}, \rr{gamPhi} and \rr{relge}. We also used
\be\label{scKx}
\fl\sinh K^*_y=\frac{b}{\sqrt{a^2-b^2}}=\sqrt\frac{1-\kp}{2\kp}\,,\qquad \cosh K^*_y=\frac{a}{\sqrt{a^2-b^2}}=
\sqrt\frac{1+\kp}{2\kp}\,.
\ee

\subsection{Ferromagnetic phase: $0<\kp<1$,  $\sqrt{1-\kp^2}<h<1$}
\label{ferro}

Let us give a general comment on the continuation of the formulas from the region $h>1$  to region $0\le h<1$.
Formally, all the formulas for matrix elements of spin operators are correct for the paramagnetic phase where $h>1$
and for the ferromagnetic phase where $0\le h<1$. But for the case $0\le h<1$ it is natural
to change the sign of $\ve(0)$ (and also of $\g(0)$) of zero-momentum excitation to be positive: $\ve(0)=1-h$.
{}From \rr{energyR}, this change of the sign of $\ve(0)$ (and of $\g(0)$) in the ferromagnetic phase
leads to a formal change between absence/presence of zero-momentum excitation
in the labelling of eigenstates. Therefore
the number of the excitations in each sector (NS and R) becomes even.
Direct calculation shows that the change of the sign of $\g(0)$ in \rr{MEsx} and \rr{MEsy}
can be absorbed to obtain formally the same formulas \rr{MEsx} and \rr{MEsy} but with
new $\g(0)$, even $L$ (the number of the excitations in R-sector) and
new $\xi=(1-{\sf k}^2)^{1/4}=((1-h^2)/\kp^2)^{1/4}$.

The explicit formulas for the matrix elements  for {\it the region of parameters} $0<\kp<1$,  $\sqrt{1-\kp^2}<h<1$ are given \cite{XY_UJP}.

\subsection{Region $\kp>1$}

{}From the relation \rr{relge} between the energies of XY-chain and Ising model excitations  it follows that the energies
$\g(p)$ of the Ising model excitations become complex and it is useful to introduce $\g(p)={\rm i}\tilde\g(p)$ such that
\[
 \sin\tilde\g(p)=\frac{\sqrt{\kp^2-1}}{\kp\sqrt{\kp^2+h^2-1}}\, \ve(p)\,.
\]
Here $\tilde\g(p)$ should be chosen to be monotonically increasing function of $p$ when $p$ runs from $0$ to $\pi$.
If $h\ge\kp^2-1$, the energy $\ve(p)$ is monotonically increasing function of $p$ with minimum at $p=0$ and maximum at $p=\pi$.
In this case $0<\tilde\g(p)\le\pi/2$.
If $0\le h<\kp^2-1$, the energy $\ve(p)$ is non-monotonical function having additional extremum (maximum) at $p=p_c$, $\cos p_c =h/(1-\kp^2)$,
$\tilde\g(p_c)=\pi/2$. In this case $0<\tilde\g(p)<\pi/2$ for $0\le p<p_c$ and  $\tilde\g(p)>\pi/2$ for $p_c<p\le \pi$.

We continue analytically the formulas  \rr{MEsx} and \rr{MEsy} and
write them in terms of $\tilde\g(p)$. All the changes in the final formulas are the following:
\[
|{}_{\rm XY}\la \Phi_0|\sx{m}|\Phi_1\ra_{\rm XY}|^2=
\frac{2\kp}{\kp+1} \cos^2\frac{\tilde\gamma_{\Phi_0}-\tilde \gamma_{\Phi_1}}{2}\, \tilde \Xi_{\Phi_0,\Phi_1}\,,
\]\[
|{}_{\rm XY}\la \Phi_0|\sy{m}|\Phi_1\ra_{\rm XY}|^2=
\frac{2\kp}{\kp-1} \sin^2\frac{\tilde\gamma_{\Phi_0}-\tilde\gamma_{\Phi_1}}{2}\, \tilde\Xi_{\Phi_0,\Phi_1}\,,
\]
where  $\tilde \Xi_{\Phi_0,\Phi_1}$  is given by  \rr{ME_BL} with substitutions
\[
\sinh\frac{\g(p)+\g(q)}{2}\to \sin\frac{\tilde\g(p)+\tilde\g(q)}{2},\quad
\frac{t_y-t_y^{-1}}{t_x-t_x^{-1}}\to \frac{\kp^2-1}{\kp\sqrt{\kp^2+h^2-1}}
\]
and $\tilde\gamma_{\Phi}$ is given by \rr{gamPhi} with $\g(p)\to \tilde\g(p)$.
Also, for $0\le h<1$, one has to take into account the modifications related to the zero mode described in Sect.~\ref{ferro}.

\subsection{Oscillatory region $0<\kp<1$, $0\le h<\sqrt{1-\kp^2}$} Similarly to the region $\kp>1$, the energies $\g(p)$
of the Ising model excitations become complex and it is useful to rewrite the matrix elements of spin operators
in terms of $\tilde \g(p)=\g(p)+{\rm i \pi}/2$:
\[
 \cosh\tilde\g(p)=\frac{\sqrt{1-\kp^2}}{\kp\sqrt{1-\kp^2-h^2}}\, \ve(p)\,.
\]
Here $\tilde\g(p)$ should be chosen to be monotonically increasing function of $p$ when $p$ runs from $0$ to $\pi$.
If $1-\kp^2\le h<\sqrt{1-\kp^2}$,
the energy $\ve(p)$ is monotonically increasing function of $p$ with minimum at $p=0$ and maximum at $p=\pi$.
In this case $\tilde\g(p)\ge 0$.
If $0\le h<1-\kp^2$, the energy $\ve(p)$ is non-monotonical function having additional extremum (minimum) at $p=p_c$, $\cos p_c =h/(1-\kp^2)$,
$\tilde\g(p_c)=0$. In this case $\tilde\g(p)<0$ for $0\le p<p_c$ and  $\tilde\g(p)>0$ for $p_c<p\le \pi$.
We continue analytically the formulas  \rr{MEsx} and \rr{MEsy} and
write them in terms of $\tilde\g(p)$. All the changes in the final formulas are the following:
\[\fl
|{}_{\rm XY}\la \Phi_0|\sx{m}|\Phi_1\ra_{\rm XY}|^2=
\frac{\kp}{2(1+\kp)} \left(e^{\frac{\tilde \gamma_{\Phi_0}-\tilde\gamma_{\Phi_1}}{2}}+(-1)^\frac{K-L}{2}
e^{-\frac{\tilde\gamma_{\Phi_0}-\tilde\gamma_{\Phi_1}}{2}}\,\right)^2 \tilde \Xi_{\Phi_0,\Phi_1}\,,
\]\[\fl
|{}_{\rm XY}\la \Phi_0|\sy{m}|\Phi_1\ra_{\rm XY}|^2=\frac{\kp}{2(1-\kp)}
 \left(e^{\frac{\tilde\gamma_{\Phi_0}-\tilde\gamma_{\Phi_1}}{2}}-(-1)^\frac{K-L}{2}
e^{-\frac{\tilde\gamma_{\Phi_0}-\tilde\gamma_{\Phi_1}}{2}}\,\right)^2 \tilde\Xi_{\Phi_0,\Phi_1}\,,
\]
where  $\tilde \Xi_{\Phi_0,\Phi_1}$  is given by  \rr{ME_BL} with $\xi=(1-{\sf k}^2)^{1/4}=((1-h^2)/\kp^2)^{1/4}$
(since it is the ferromagnetic phase) and with the repacements
\[
\sinh\frac{\g(p)+\g(q)}{2}\to \cosh\frac{\tilde\g(p)+\tilde\g(q)}{2},\quad
\frac{t_y-t_y^{-1}}{t_x-t_x^{-1}}\to \frac{1-\kp^2}{\kp\sqrt{1-\kp^2-h^2}}\,,
\]
$\tilde\gamma_{\Phi}$ is given by \rr{gamPhi} with $\g(p)\to \tilde\g(p)$.
Also one has to take into account the modifications related to the zero mode described in Sect.~\ref{ferro}.

\subsection{Other values of parameters}

\subsubsection{Quantum Ising chain: $\kp=1$.}
In the case of the quantum Ising chain ($\kp=1$) the formula for the matrix elements of spin operator $\sx{m}$
can be derived by a limiting procedure. The final formula was derived in \cite{gipst2,IST} and
it is expressed in terms of the energies of excitations $\ve(q)$. In the case of general
XY-chain we were not able to find an analogous formula for the matrix elements in terms of $\ve(q)$.

\subsubsection{Boundary of oscillator region: $\kp^2+h^2=1$.}
One of the peculiarities of XY-chain when the parameters belong to the curve $\kp^2+h^2=1$ 
is that the ground states $|\rangle^{\rm NS}_{\rm XY}$ and
$|\rangle_{\rm XY}^{\rm R}$ are degenerate and each of them can be presented as a sum of two pure tensors
\cite{KTM}. In \cite{BarouchMcCoy}, for such values of parameters, the two-point correlation function was found explicitly.
Here we give some comments on the matrix elements of spin operators.
They  can be derived from the general formulas
for the region $0<\kp<1$,  $\sqrt{1-\kp^2}<h<1$ by a limiting procedure.
From \rr{relge}, denoting $\zeta=\sqrt{\kp^2+h^2-1}$ we get in the limit $\zeta\to 0$
\[
e^{\g(p)}=\frac{2h}{\zeta\sqrt{1-h^2}}\,\ve(p)\,,\qquad \ve(p)=1-h\cos p\,,
\]\[
\sinh \frac{\g(p)+\g(q)}{2} = \frac{e^{(\g(p)+\g(q))/2}}{2}\,, \qquad \xi= 1\,, \qquad \xi_T=1\,.
\]
Using these formulas it is easy to take the limit $\zeta\to 0$ in the general formulas for the matrix elements.
We get, in particular, that the matrix elements are non-zero if and only if $K=L$ or $K-L=\pm 2$.

\section{Asymptotics of form-factors in the limit of infinite chain}

In this section we analyze the asymptotics of different parts of form-factors in the limit of infinite
length ($n\to \infty$) of XY-chain. For this end it is helpful to use the following integral representations
for different parts of form-factors at finite $n$ \cite{BL2}. For
\[{
\Lambda^{-1}=\frac{1}{2} \left(\sum_{\qu\in {\rm NS}}\g(\qu)-\sum_{\pu\in {\rm R}}\g(\pu)\right)\,,\qquad}
e^{\eta(q)}=\frac{\prod_{p\in {\rm NS}}\left(1-e^{-\g(q)-\g(p)}\right)}{\prod_{p\in {\rm R}}\left(1-e^{-\g(q)-\g(p)}\right)}
\]
and $\xi_T$ (see \rr{xiT})  we have
\[
\Lambda^{-1}=\frac{1}{\pi}\int_0^\pi dp\; \log \coth \frac{n \bar\g(p)}{2}\,,
\]\[
\eta(q)=\frac{1}{\pi}\int_0^\pi dp\; \frac{\cos p-e^{-\g(q)}}{\cosh \g(q)-\cos p} \log \coth \frac{n \bar\g(p)}{2}\,,
\]\[
\xi_T=\frac{n^2}{2\pi^2}
\int_0^\pi\int_0^\pi \frac{dp\; dq\;\bar\g'(p)\bar\g'(q)}{\sinh n \bar\g(p)\;\sinh n \bar\g(p)}
\log\left| \frac {\sin ((p+q)/2)}{\sin ((p-q)/2)}\right|\,.
\]

Let us show that  $\Lambda^{-1}\to 0$ if $n\to \infty$. In fact we have the following asymptotics
\[
\Lambda^{-1}\to \frac{2}{\pi}\int_0^\pi dp\; e^{-n \bar\g(p)}=
\frac{2}{\pi}\int_0^\pi dp\; e^{-n (\bar\g(0)+\bar\g''(0)p^2/2+\cdots)}
\]\[\qquad\simeq
\frac{e^{-n \bar\g(0)}}{\pi}\int_{-\infty}^{\infty} dp\; e^{-n \bar\g''(0)p^2/2}=
e^{-n \bar\g(0)}\sqrt\frac{2}{n \pi \bar\g''(0)}\to 0\,,
\]
where we used the fact that $\bar\g(p)>0$ (there is a gap in the spectrum for the non-critical parameters).
Similarly  we get $\eta(q)\to 0$, $\xi_T\to 1$ at $n\to \infty$. For the derivation of these formulas together with the more
precise asymptotics at $n\to \infty$, see \cite{Bugrij}.
Another way to get the asymptotics  in the limit of infinite length of the chain is given in \cite{XY_UJP}.

In the limit of infinite XY-chain the formulas \rr{MEsx}, \rr{MEsy}
for the matrix elements of spin operators between the eigenstates
$|\Phi_0\ra_{\rm XY}=|\qu_1, \qu_2,\ldots,\qu_K\ra_{\rm XY}^{\rm NS}$ from the NS-sector
and $|\Phi_1\ra_{\rm XY}=|\pu_1, \pu_2,\ldots,\pu_L\ra_{\rm XY}^{\rm R}$ from the R-sector,
and  \rr{ME_BL} become
\be\label{MEsx_inf}
\fl |{}_{\rm XY}\la \Phi_0|\sigma^x_{m}|\Phi_1\ra_{\rm XY}|^2=
\Xi_{\Phi_0,\Phi_1}\, {\frac{2\kp}{1+\kp}} \,  \cosh^2 \frac {\sum_{k=1}^K  \g(\qu_k)-\sum_{l=1}^L \g(\pu_l)}{2}  \,,
\ee\be\label{MEsy_inf}
\fl |{}_{\rm XY}\la \Phi_0|\sigma^y_{m}|\Phi_1\ra_{\rm XY}|^2=
\Xi_{\Phi_0,\Phi_1}\, \frac{2\kp}{1-\kp} \,  \sinh^2 \frac {\sum_{k=1}^K  \g(\qu_k)-\sum_{l=1}^L \g(\pu_l)}{2}\,,
\ee\[\fl
\Xi_{\Phi_0,\Phi_1}=\;\xi\;  \prod_{k=1}^K \frac{1} {n \sinh \g(\qu_k)}\;\;
\prod_{l=1}^L \;\frac{1} {n \sinh \g(\pu_l)}\cdot
\left(\frac{t_y-t_y^{-1}}{t_x-t_x^{-1}}\right)^{\!\!(K-L)^2/2}
\]\be\times\;
\prod_{k<k'}^K \frac{\sin^2\frac{\qu_k-\qu_{k'}}{2}} {\sinh^2 \frac{\g(\qu_k)+\g(\qu_{k'})}{2}}
\;\;\prod_{l<l'}^L \frac{\sin^2\frac{\pu_l-\pu_{l'}}{2}} {\sinh^2 \frac{\g(\pu_l)+\g(\pu_{l'})}{2}}
\prod_{1\le k \le K \atop 1\le l \le L}
\frac {\sinh^2 \frac{\g(\qu_k)+\g(\pu_l)}{2}} {\sin^2\frac{\qu_k-\pu_l}{2}}\,, \label{ME_BL_inf}
\ee
where the relation \rr{relge} between $\g(q)$ and the energy of excitation $\ve(q)$ in XY-chain and relation
\rr{tytx} are also to be used.

In the ferromagnetic phase ($0\le h<1$) after an appropriate modification as explained in Sect.~\ref{ferro},
the thermodynamic limit formulas \rr{MEsx_inf} and \rr{MEsy_inf} at $K=L=0$ allow to re-obtain \cite{XY_UJP}
formulas for the spontaneous magnetization found in \cite{Suzuki}.
Indeed, in the thermodynamic limit $n\to\infty$,
 the energies of vacuum states $|\Phi_0\ra_{\rm XY}=|\rangle^{\rm NS}_{\rm XY}$ and
$|\Phi_1\ra_{\rm XY}=|\rangle_{\rm XY}^{\rm R}$ of different sectors asymptotically coincide giving
the degeneration of the ground state of the Hamiltonian. It leads to a non-zero value of vacuum Bogoliubov quasiaverage
of $\sigma^{\rm x}$ which is spontaneous magnetization:
\[\fl
\la\sigma^{\rm x,y}\ra_{\rm XY}={}_{\rm XY}^{\rm R}\la |\sigma^{\rm x,y}|\ra_{\rm XY}^{\rm NS}\,,\qquad
\la\sx{}\ra_{\rm XY}=\sqrt{2}\left(\frac{\kp^2(1-h^2)}{(1+\kp)^4}\right)^{1/8},\qquad
\la\sy{}\ra_{\rm XY}=0\,.
\]
The matrix element ${}_{\rm XY}^{\rm R}\la |\sigma^{\rm x}|\ra_{\rm XY}^{\rm NS}$ give spontaneous magnetization
because it appears as $0$-particle contribution from R-sector in the long-distance expansion of two-point correlation function
$\la \sx{0}\sx{d}\ra$ in the ferromagnetic phase.

\section{Asymptotics of the two-point correlation function of the infinite XY-chain}

We are interested in the asymptotics of the two-point correlation function when the distance $d$ between the correlating spins $\sx{0}$
and $\sx{d}$ is large (and, of course, the length $n$ is much larger). We show how to re-derive the formula
from \cite{BarouchMcCoy} without the use of Szeg\"o theorem and Wiener-Hopf method for finding
the asymptotics of Toeplitz determinant.

For definiteness we consider the paramagnetic phase.
In this case the intermediate states between two spin operators are odd-number particle states
from R-sector.
Let us show that the main contribution is due to $1$-particle states
(the contribution of many particle states is exponentially suppressed for large $d$).
We have\footnote{In this section all the eigenstates of XY-chain Hamiltonian are labeled only by NS or R depending on the sector.}
\[\fl
\la \sx{0}\sx{d}\ra= \sum_{p\in {\rm R}} {}_{\rm NS}\la|\sx{0}| p\ra_{\rm R} \cdot
{}_{\rm R}\la p |\sx{d}|\ra_{\rm NS}\]
\[ +\sum_{\{p_1<p_2<p_3\}\in {\rm R}} {}_{\rm NS}\la|\sx{0}| p_1,p_2,p_3\ra_{\rm R} \cdot
{}_{\rm R}\la p_1,p_2,p_3 |\sx{d}|\ra_{\rm NS}+\cdots
\]\[\fl
=\sum_{p\in {\rm R}} e^{{\rm i}p d}
|{}_{\rm NS}\la|\sx{0}| p\ra_{\rm R}|^2+
\sum_{\{p_1<p_2<p_3\}\in {\rm R}} e^{{\rm i}(p_1+p_2+p_3) d}|{}_{\rm NS}\la|\sx{0}| p_1,p_2,p_3\ra_{\rm R}|^2+\cdots\,,
\]
where $|\ra_{\rm NS}$ is the vacuum state.
The idea is to make a transformation which corresponds to change of the direction of evolution to the transverse direction
in the lattice formulation of the Ising model.
Let us estimate the $L$-particle contribution (in the limit $n\to \infty$ we use integrals instead of sums):
\[\fl
\cosh^2 K^*_y\,\sum_{\{p_1<\cdots<p_L\}\in {\rm R}} e^{{\rm i}(p_1+\cdots+p_L) d}|{}_{\rm NS}\la|\sx{0}| p_1,\ldots,p_L\ra_{\rm R}|^2=
\xi\; \left(\frac{t_y-t_y^{-1}}{t_x-t_x^{-1}}\right)^{L^2/2}
\]\be\fl\quad
\times \frac{1}{(2\pi)^L L!}\int_{-\pi}^\pi \cdots\int_{-\pi}^\pi  \prod_{l=1}^L \;\frac{d\pu_l} {\sinh \g(\pu_l)}
\, e^{{\rm i}d \sum_{l=1}^L \pu_l}
\cosh^2 \frac {\sum_{l=1}^L \g(\pu_l)}{2}
\prod_{l<l'}^L \frac{\sin^2\frac{\pu_l-\pu_{l'}}{2}} {\sinh^2 \frac{\g(\pu_l)+\g(\pu_{l'})}{2}}\,. \label{Lpart}
\ee
Now we change the variables in the integrals from the momenta $\{\pu_l\}$
to parameters $\{u_l\}$ on the elliptic curve by means of relations from Sect.~\ref{elparam} and from \cite{Palmer}
\[\fl\qquad
du_l=\frac{t_y^{-1}-t_y}{4} \frac{dp_l}{\sinh\g(p_l)}\,,\qquad
\frac{\sin\frac{\pu_l-\pu_{l'}}{2}} {\sinh \frac{\g(\pu_l)+\g(\pu_{l'})}{2}}=
\frac{t_x^{-1} - t_x}{2\,{\sf k}}\, \, \mbox{sn}(u_l - u_{l'}, {\sf k}^{-1})\,.
\]
Thus the $L$-particle contribution is
\[\fl
\xi\; \left(\frac{t_y-t_y^{-1}}{t_x-t_x^{-1}}\right)^{L^2/2} \frac{1}{(2\pi)^L L!}\frac{4^{L-1}}{(t_y^{-1}-t_y)^L}
\left(\frac{t_x^{-1} - t_x}{2\, {\sf k}}\right)^{L(L-1)}
\int_{0}^{2K} \cdots \int_{0}^{2K} du_1\ldots du_L
\]\[\fl\times
\prod_{l=1}^L z(u_l+{\rm i}K'/2)^d \left(\prod_l\lm(u_l+{\rm i}K'/2)^{1/2}+\prod_l \lm(u_l+{\rm i}K'/2)^{-1/2}\right)^2
\prod_{l<l'}^L \, \mbox{sn}^2(u_l - u_{l'}, {\sf k}^{-1})\,.
\]
Shifting the contours of the integrations by $-{\rm i} K'/2$ we get
\[\fl
\xi\; \left(\frac{t_y-t_y^{-1}}{t_x-t_x^{-1}}\right)^{L^2/2} \frac{1}{(2\pi)^L L!}\frac{4^{L-1}}{(t_y^{-1}-t_y)^L}
\left(\frac{t_x^{-1} - t_x}{2\,{\sf k}}\right)^{L(L-1)}
\int_{0}^{2K} \cdots \int_{0}^{2K} du_1\ldots du_L
\]\[\fl\qquad\times
\prod_{l=1}^L z(u_l)^d \left(\prod_l\lm(u_l)^{1/2}+\prod_l \lm(u_l)^{-1/2}\right)^2
\prod_{l<l'}^L \, \mbox{sn}^2(u_l - u_{l'}, {\sf k}^{-1})\,.
\]
With the use of the relations $\lm(u_l)=e^{{\rm i} \bar p_l}$, $z(u_l)=e^{-\bar\g(\bar p_l)}$,
\[\fl\qquad
du_l=\frac{t_x^{-1}-t_x}{4} \frac{d\bar p_l}{\sinh\bar\g(\bar p_l)}\,,\qquad
\frac{\sin\frac{\bar\pu_l-\bar\pu_{l'}}{2}} {\sinh \frac{\bar\g(\bar\pu_l)+\bar\g(\bar\pu_{l'})}{2}}=
\frac{t_y^{-1} - t_y}{2\,{\sf k}}\, \mbox{sn}(u_l - u_{l'}, {\sf k}^{-1})\,,
\]
the $L$-particle contribution to the two-point correlation function can be rewritten in the terms of energies and momenta
corresponding to the transverse direction of the Ising lattice:
\[
\xi\; \left(\frac{t_x-t_x^{-1}}{t_y-t_y^{-1}}\right)^{L^2/2} \frac{1}{(2\pi)^L L!}
\int_{-\pi}^{\pi} \cdots \int_{-\pi}^{\pi} \frac{dp_1}{\sinh \bar\g(p_1)}\ldots \frac{dp_L}{\sinh \bar\g(p_L)}
\]\be\qquad\times
 e^{-d \sum_{l=1}^L \bar\g(\pu_l)} \cos^2 \frac {\sum_{l=1}^L \pu_l}{2}
\prod_{l<l'}^L \frac{\sin^2\frac{\pu_l-\pu_{l'}}{2}} {\sinh^2 \frac{\bar\g(\pu_l)+\bar\g(\pu_{l'})}{2}}\,. \label{Lpart_tr}
\ee
In the limit $d\to\infty$ the main contribution of these integrals is given by the neighborhood of zero momenta, where
$\bar\g(p)$ has a minimum. When the parameters of the model are non-critical we have $0<\exp(-\bar\g(0))=a^2<1$
and the asymptotics of $L$-particle contribution to the two-point correlation function is proportional to
$e^{-dL\bar\g(0)}$. Thus the $1$-particle contribution dominates in the paramagnetic phase.

{}From \rr{Lpart} for the $1$-particle contribution we have
\[
\cosh^2 K^*_y\, \la \sx{0}\sx{d}\ra\simeq
\frac{\xi }{2\pi}\left(\frac{t_y-t_y^{-1}}{t_x-t_x^{-1}}\right)^{\!\!1/2}
\int_{-\pi}^{\pi} dp\, e^{{\rm i}p d} \coth \frac{\g(p)}{2}\,.
\]
Now we take into account \rr{drBBS}, \rr{erIsBBS}
in order to rewrite the formula in terms of $a$ and $b$. With $z=e^{{\rm i}p}$ we have
\[
\coth \frac{\g(p)}{2}=\frac{a}{b} s_p=
\frac{a}{b} \left(\frac{(b^2-z)(b^2-z^{-1}}{(a^2-z)(a^2-z^{-1})}\right)^{1/2}=
\sqrt\frac{(b^2-z)(b^{-2}-z)}{(a^2-z)(a^{-2}-z)}\,.
\]
We rewrite the integral as a contour integral:
\[
I(d)=\frac{1}{2\pi}\int_{-\pi}^\pi dp\;e^{{\rm i}p d} \coth \frac{\g(p)}{2}=
\frac{1}{2\pi{\rm i}}\oint_{|z|=1}dz\; z^{d-1} \sqrt\frac{(b^2-z)(b^{-2}-z)}{(a^2-z)(a^{-2}-z)}\,.
\]
To fix a branch of function under the integral we make two cuts: from $b^2$ to $a^2$ ($0<b^2<a^2<1$)
and from $a^{-2}$ to $b^{-2}$ ($1<a^{-2}<b^{-2}$).
We move the contour to be along the cut from $b^2$ to $a^2$:
\[
I(d)=\frac{1}{\pi}\int_{b^2}^{a^2} ds\;s^{d-1} \sqrt\frac{(s-b^2)(b^{-2}-s)}{(a^2-s)(a^{-2}-s)}
\]
Since we consider the limit $d\to \infty$, the the main contribution to the integral
is given by  the neighborhood of $a^2$ due to the factor $s^{d-1}$.
We change the variable of integration to $t=a^2-s$:
\[
I(d)=\frac{1}{\pi}\int_{0}^{a^2-b^2} \frac{ dt}{\sqrt{t}} \;(a^2-t)^{d-1}
\sqrt\frac{(a^2-b^2-t)(b^{-2}-a^2+t)}{a^{-2}-a^2+t}
\]
and use the expansion at $d\to\infty$ and $t\sim 0$
\[
(a^2-t)^{d-1}=e^{(d-1)(\log a^2+\log(1-t/a^2)}\sim a^{2(d-1)} e^{-(d-1)t/a^2}
\]
to get the leading asymptotics by steepest descent method:
\[
I(d)\sim a^{2(d-1)}  (d-1)^{-1/2} \sqrt\frac{a^2(a^2-b^2)(b^{-2}-a^2)}{\pi(a^{-2}-a^2)}\,.
\]
Subsequent terms of the asymptotics for the $1$-particle contributions are given by (4.22) of \cite{BarouchMcCoy}.
Finally the leading term of asymptotics of the two-point correlation function is
\be\label{asy_2p}\fl
\la \sx{0}\sx{d}\ra\sim
\frac{\xi}{2}\left(\frac{t_y-t_y^{-1}}{t_x-t_x^{-1}}\right)^{\!\!1/2} \frac{I(d)}{\cosh^2 K^*_y}
\sim {a^{2d}} { d^{-1/2}} \pi^{-1/2}\cdot
\left(\frac{(1-b^2)(1-b^2/a^2)^2}{ 1-a^4}\right)^{1/4}\,,
\ee
where we used \rr{xi},\rr{tytx} and \rr{scKx}.
It coincides up to factor $(-1)^d/4$ with (4.25) of \cite{BarouchMcCoy} after identification $\lm_1=b^{-2}$, $\lm_2=a^2$.
The origin of this discrepancy factor is the following: In \cite{BarouchMcCoy}
the spin operators are $1/2$ of Pauli matrices. It gives factor $1/4$ for the two-point correlation functions.
Also in \cite{BarouchMcCoy}, the coefficients at $\sx{k} \sx{k+1}$ and $\sy{k} \sy{k+1}$ in the Hamiltonian
are positive. In this paper they are negative. To change the signs we make
automorphism of the algebra of spin operators:
$\sx{k}\to (-1)^k \sx{k}$, $\sy{k}\to (-1)^k \sy{k}$, $\sz{k}\to \sz{k}$. This map changes the signs of the mentioned coefficients
and leads to the factor $(-1)^d$ in the two-point correlation function.
To rewrite the correlation function \rr{asy_2p} in terms of $h$ and $\kp$ one needs \rr{abhkp}.

Another way to calculate the two-point correlation function is to use integral representation for the $L$-particle
contribution \rr{Lpart_tr} in terms of energies in transverse
direction on the Ising lattice:
\[\fl
\cosh^2 K^*_y\,\cdot \la \sx{0}\sx{d}\ra\sim \frac{\xi}{2\pi}\,
\frac{t_x^{-1}-t_x}{t_y^{-1}-t_y} \int_{-\pi}^\pi dp\;  \frac{e^{-d\bar\g(p)}} {\sinh \bar\g(p)}\cos^2 \frac {p}{2}
\]\[\fl\simeq \frac{\xi}{2\pi}\,
\frac{t_x^{-1}-t_x}{t_y^{-1}-t_y} \cdot\frac{1}{\sinh \bar\g(0)}\int_{-\pi}^\pi dp\;  e^{-d(\bar\g(0)+\bar\g''(0)p^2/2)}\simeq
\frac{\xi}{2\pi} \, \frac{t_x^{-1}-t_x}{t_y^{-1}-t_y} \cdot\frac{e^{-d\bar\g(0)}}{d^{1/2} \sinh \bar\g(0)}\sqrt\frac{2\pi}{\bar\g''(0)}\,.
\]
We use also \rr{drIsdual} and \rr{tytx}
\[\fl
\bar\g''(0)=\frac{t_x^{-1}-t_x}{t_y^{-1}-t_y} \cdot\frac{1}{ \sinh \bar\g(0)}\,,\qquad
e^{-\bar\g(0)}=a^2\,,\qquad
\sinh \bar\g(0)=\frac{a^{-2}-a^2}{2}
\]
to obtain by steepest descent method the same asymptotics of the two-point correlation function \rr{asy_2p}.

To obtain more precise asymptotics, we change the variable of integration $s=\exp(-\bar\g(p))$,
$-ds/s= d\bar\g(p)$ and use the dispersion relation \rr{drIsdual} to get
\[\frac {dp}{ \sinh\bar\g(p)}=
\frac{d\bar \g}{\sin p}\cdot
\frac{t_y^{-1}-t_y}{t_x^{-1}-t_x} =-\frac{ds}{s\,\sin p}\cdot
\frac{t_y^{-1}-t_y}{t_x^{-1}-t_x}\,.
\]
It gives
\[
\frac{t_x^{-1}-t_x}{t_y^{-1}-t_y} \int_{-\pi}^\pi dp\;  \frac{e^{-\bar\g(p)d}} {\sinh \bar\g(p)}\cos^2 \frac {p}{2}=
\int_{b^2}^{a^2} ds \; s^{d-1} \cot \frac {p}{2}
\]
{}From \rr{drIsdual} we have
\[
\cot \frac {p}{2}=\sqrt{\frac{(s-b^2)(b^{-2}-s)}{(a^2-s)(a^{-2}-s)}}\,.
\]
Substituting this expression into the integral we get the formula coinciding with formula (4.21) from \cite{BarouchMcCoy}.
It can be expanded as described there to obtain the subsequent terms of the asymptotics.

\section{Discussion}

Using the results of \cite{gipst2} on the separation of variables method for the Baxter--Bazhanov--Stroganov model,
in this paper we derived matrix elements (form-factors) of spin operators $\sx{}$ and $\sy{}$ between the eigenvectors of the Hamiltonian
\rr{HamXYh} of finite (length $n$) XY-chain. The final formulas are \rr{MEsx}, \rr{MEsy} and \rr{ME_BL}.
In the limit of infinite chain the formulas are reduced to \rr{MEsx_inf}, \rr{MEsy_inf} and \rr{ME_BL_inf}
and allow to re-derive the asymptotics of correlation function $\la\sx{0}\sx{d}\ra$ at $d\to \infty$ in a simple way.
It is interesting to derive analytically the long-time and long-distance asymptotic behavior of the $\sx{}$ spin correlations at finite temperature
observed recently in \cite{KrSt} by numeric analysis.

Usually the standard efficient way to study the Ising model and XY-chain is to use the language of fermions.
Recently the factorized formulas for the matrix elements of spin operators were re-derived for the quantum Ising chain \cite{IST},
for the Ising model on two-dimensional lattice \cite{IL}, for general free fermion model on two-dimensional lattice and quantum XY-chain
\cite{ILff} using algebra of fermion operators.

\ack
The author thanks A.~Bugrij, A.~Kl\"umper, O.~Lisovyy and V.~Shadura  for helpful discussions.
The work of the author was partially supported
by the Program of Fundamental Research of the Physics and
Astronomy Division of the NAS of Ukraine, the Ukrainian-Russian FRSF-RFBR project,
by French-Ukrainian  joint project PICS of CNRS and NAS of Ukraine.

\subsection*{References}
\bibliographystyle{amsplain}

\end{document}